\documentclass[12pt]{iopart}

\usepackage{iopams}
\expandafter\let\csname equation*\endcsname\relax
\expandafter\let\csname endequation*\endcsname\relax 
\usepackage{amsmath,amssymb,amsfonts}
\usepackage{algorithmic}
\usepackage{textcomp}
\usepackage{subfigure}
\usepackage{multirow}
\usepackage{xcolor}
\usepackage{booktabs}
\usepackage{color, colortbl}
\usepackage{amssymb}
\usepackage{epstopdf}
\usepackage{cite}
\usepackage{url}
\usepackage{hyperref}
\usepackage{graphicx}
\usepackage{harvard}  
\newcommand\keywords[1]{\textbf{Keywords}: #1}
\begin{document}

\title[Optimizing Medical Image Segmentation with Advanced Decoder Design]{Optimizing Medical Image Segmentation with Advanced Decoder Design}

\author{Weibin Yang\textsuperscript{1}, Zhiqi Dong\textsuperscript{1}, Mingyuan Xu\textsuperscript{1}, Longwei Xu\textsuperscript{1}, Dehua Geng\textsuperscript{1}, Yusong Li\textsuperscript{1}, Pengwei Wang\textsuperscript{1,*}}
\address{\textsuperscript{1} School of Information Science and Engineering, Shandong University, Tsingtao, 266237, China}
\address{\textsuperscript{*}
Corresponding Author }

\ead{wangpw@sdu.edu.cn}
\vspace{10pt}
\begin{indented}
\item[]October 2024
\end{indented}

\begin{abstract}

U-Net is widely used in medical image segmentation due to its simple and flexible architecture design. To address the challenges of scale and complexity in medical tasks, several variants of U-Net have been proposed. In particular, methods based on Vision Transformer (ViT), represented by Swin UNETR, have gained widespread attention in recent years. However, these improvements often focus on the encoder, overlooking the crucial role of the decoder in optimizing segmentation details. This design imbalance limits the potential for further enhancing segmentation performance. To address this issue, we analyze the roles of various decoder components, including upsampling method, skip connection, and feature extraction module, as well as the shortcomings of existing methods. Consequently, we propose Swin DER (i.e., Swin UNETR Decoder Enhanced and Refined) by specifically optimizing the design of these three components. Swin DER performs upsampling using learnable interpolation algorithm called offset coordinate neighborhood weighted up sampling (Onsampling) and replaces traditional skip connection with spatial-channel parallel attention gate (SCP AG). Additionally, Swin DER introduces deformable convolution along with attention mechanism in the feature extraction module of the decoder. Our model design achieves excellent results, surpassing other state-of-the-art methods on both the Synapse and the MSD brain tumor segmentation task.

Code is available at: 
\href{https://github.com/WillBeanYang/Swin-DER}{\textit{https://github.com/WillBeanYang/Swin-DER}}
\end{abstract}


%
%
%
%
%
\section{Introduction}
\label{sec:introduction}
Medical image segmentation aims to utilize computer technology to extract boundaries of anatomical or pathological features from medical images\cite{overview1}. Precise segmentation results can provide reliable volume and structural information of the target structure, thereby assisting clinicians in making accurate diagnostic and therapeutic decisions\cite{overview4,decathlon,wasserthal2023totalsegmentator}. As an emerging biomedical image processing technique, medical image segmentation offers crucial support for intelligent diagnosis and precise medical care\cite{overview2}.

With the advancement of artificial intelligence, especially deep learning, medical image segmentation techniques based on deep convolutional neural networks (CNNs) have achieved noteworthy outcomes\cite{overview3}. Among these, U-Net \cite{UNet} has become one of the most extensively utilized architectures in medical image segmentation due to its optimized module design and flexibility\cite{overview5}. U-Net employs an encoder-decoder architecture, in which the encoder progressively extracts and compresses features, while the decoder is responsible for restoring spatial resolution and optimizing segmentation details\cite{overview6}. Despite the powerful representational learning capability of CNN-based methods, the locality of convolution operations limits the network's ability to learn long-distance dependencies\cite{overview7}.

Transformer\cite{transformer} can encode long-distance dependencies and learn efficient feature representations. In the field of computer vision, the Vision Transformer (ViT)\cite{vit} slices images into patches, encoding them into a 1D sequence and using self-attention mechanism to capture the complex relationships between these patches. This effectively addresses the shortcomings of convolutional operations. Currently, many researchers are attempting to integrate the ViT into the improvement efforts of U-Net\cite{transunet,unetr,swinunet}, with Swin UNETR\cite{swinunetr} being the most prominent among them. Swin UNETR is distinguished by its incorporation of two sets of encoders that harness the individual strengths of both CNNs and transformers. This design grants it a powerful feature extraction capability, enabling it to achieve competitive segmentation performance across different datasets.

However, whether based on CNNs or Transformers, most current U-Net variants confine their improvement strategies to building more complex encoder, whereas leaving the decoder unchanged or adopting a simple symmetric structure. This imbalance in design is particularly evident in Swin UNETR, which has two sets of encoders while the decoder consists of simple Residual Blocks. We believe that there is still considerable room for improvement in the current decoder design. Firstly, commonly used upsampling methods often suffer from issues, such as checkerboard artifacts in transpose convolutions.
Secondly, there is often a significant semantic gap between the encoder and decoder features that are simply merged through skip connections, which can negatively impact segmentation performance. 
Additionally, for objects with complex contours,  convolution and transformer may struggle to precisely capture their boundary and shape features, thus affecting the ability of decoder to optimize segmentation details.

In response to these issues, we propose Swin DER (i.e., \textbf{Swin} UNETR \textbf{D}ecoder \textbf{E}nhanced and \textbf{R}efined), which focuses on optimizing the decoder part of Swin UNETR. Our specific contributions are as follows:

\begin{itemize}
  \item[$\bullet$] 
We propose a novel upsampling algorithm, \textbf{O}ffset coordinate \textbf{n}eighborhood weighted up\textbf{sampling} (\textbf{Onsampling}), which optimizes interpolation algorithms using learnable position offsets and pixel weights.
  \item[$\bullet$] 
Introducing spatial and channel attention gates in parallel within skip connections to eliminate irrelevant and noisy responses in the skip connections.
  \item[$\bullet$] 
The decoder block is modified to a Deformable Squeeze-and-Attention (DSA) Block, utilizing deformable convolutions to further learn detailed features and focusing on important parts of the feature map through attention mechanisms.
\end{itemize}

Our model design achieves excellent results, surpassing other state-of-the-art methods on both the Synapse dataset\cite{BTCV} and brain tumor segmentation task in Medical Segmentation Decathlon (MSD)\cite{decathlon}.

\section{Related work}
\label{sec:relatedwork}
In this section, we review several optimization efforts aimed at the decoder components of the U-Net, including upsampling, skip connection, and feature extraction block. We also analyze the limitations of existing works.

\subsection{Upsampling}
Feature upsampling is one of the most critical operations in the decoder side of the U-Net architecture. Through upsampling, the U-Net progressively restores the resolution of feature maps to match the high-resolution supervision. Interpolation algorithms, such as nearest neighbor and bilinear interpolation, are the most commonly used methods for upsampling\cite{patil2013medical3}. These methods utilize the spatial distances between pixels to guide the upsampling process. However, due to following fixed rules to interpolate upsampled values, these algorithms fail to accurately capture the rich semantic information in medical images which are often characterized by diverse shapes and complex structures. To increase flexibility, some algorithms introduce learnable parameters in the upsampling process. For instance, transposed convolution\cite{transconv} achieves upsampling by convolving the input feature map pixels with a convolutional kernel and adjusting the size of the output feature map according to the stride and padding parameters. However, when the stride and kernel size are not appropriately matched, transposed convolution can lead to checkerboard artifacts\cite{checkerboard}. W. Shi et al. proposed sub-pixel convolution\cite{patil2013medical5}, which increases the depth of feature map channels through convolution operations and then moves the channel pixels to the spatial dimension to increase the feature map resolution. However, sub-pixel convolution is commonly used in super-resolution reconstruction tasks and does not perform well in semantic segmentation tasks. The literature \cite{shi2016deconvolution} also indicates that sub-pixel convolution and transposed convolution essentially employ the same approach to processing low-resolution feature maps. 

Unlike these learnable upsampling methods weigh pixels through convolution, we propose Onsampling, which achieves dynamic upsampling by learning offsets and neighborhood weights, effectively avoiding these issues.

\subsection{Skip connection}
Skip connections integrate deep semantic information with shallow local information, allowing the decoder to learn segmentation details more effectivelly\cite{skipconnection1}. Moreover, skip connections also aid in the better convergence of the network\cite{skipconnection2}. Zhou et al.\cite{UNet++}introduced UNet++, which incorporates dense skip connections, allowing features from different levels to be aggregated instead of being confined to the same level of encoders and decoders. Huang et al.\cite{UNet3+} believed that UNet++ still did not fully exploit full-scale features. In their architecture, UNet 3+, they further introduced full-scale skip connections, connecting each decoder with all encoders and all preceding decoders. Xiang et al.\cite{BiO-Net} added backward skip connections on the basis of forward skip connections, transferring the features of decoder to encoder at the same level. These works increase the number of skip connections to make full use of features at each stage of the network, yet they neglect that the fusion of two sets of features with a large semantic gap could interfere with the learning process of the network\cite{multiresunet}. To eliminate irrelevant and noisy responses in skip connections, Okey et al. proposed Attention U-Net \cite{Attention_UNet}, which employs attention gate (AG) to suppress encoder features that are irrelevant to the decoder features, reducing the semantic gap between the encoder and decoder features. However, AG merely adjusts the importance of encoder features in the spatial dimension, neglecting the semantic discrepancy between encoder and decoder across the channel dimension. To address this, we propose the spatial-channel parallel attention gate (SCP AG), which employs channel-wise attention gate to assign distinct spatial weight maps to each channel of encoder features.

\subsection{Feature extraction block}
After the decoder receives features concatenated via skip connection, it further merges these features through its feature extraction block, namely the decoder block, and gradually refines the details of the features. Typically, the decoder module consists of two consecutive convolution layers. Drozdzal et al.\cite{skipconnection1} proposed the Residual U-Net, employing residual blocks as the feature extraction blocks, which enabled the network to converge faster. Inspired by Inception blocks, Ibtehaz et al.\cite{multiresunet} introduced the MultiRes block, adapting the Inception blocks concept by representing 5$\times$5 and 7$\times$7 convolutions as sequences of two and three 3$\times$3 convolutions, respectively, and incorporating a residual connection with 1$\times$1 convolution. With the rise of ViT in the field of computer vision, an increasing number of methods have started to utilize ViT as their feature extraction block\cite{unetr,nnformer}. However, convolutions can only extract regular local features of objects, while ViT focuses more on capturing global features. Using them as feature extractors in decoders can not accurately capture the boundaries and detailed features of objects. The advent of deformable convolution\cite{deformable} provides us with a new perspective. In deformable convolution, each position of the convolution kernel can dynamically offset based on the input data. These offsets are learned by the network, allowing the convolution kernel to adaptively adjust its shape to better match the local detail features of the input image. Consequently, we designed the Deformable Squeeze-and-Attention (DSA) Block as the feature extraction block for the decoder, replacing the second convolution layer in the Squeeze-and-Attention (SA) Block with deformable convolution.

\section{Method}

\subsection{Overview}
\begin{figure}[!t]
    \centering
    \includegraphics[width=\columnwidth]{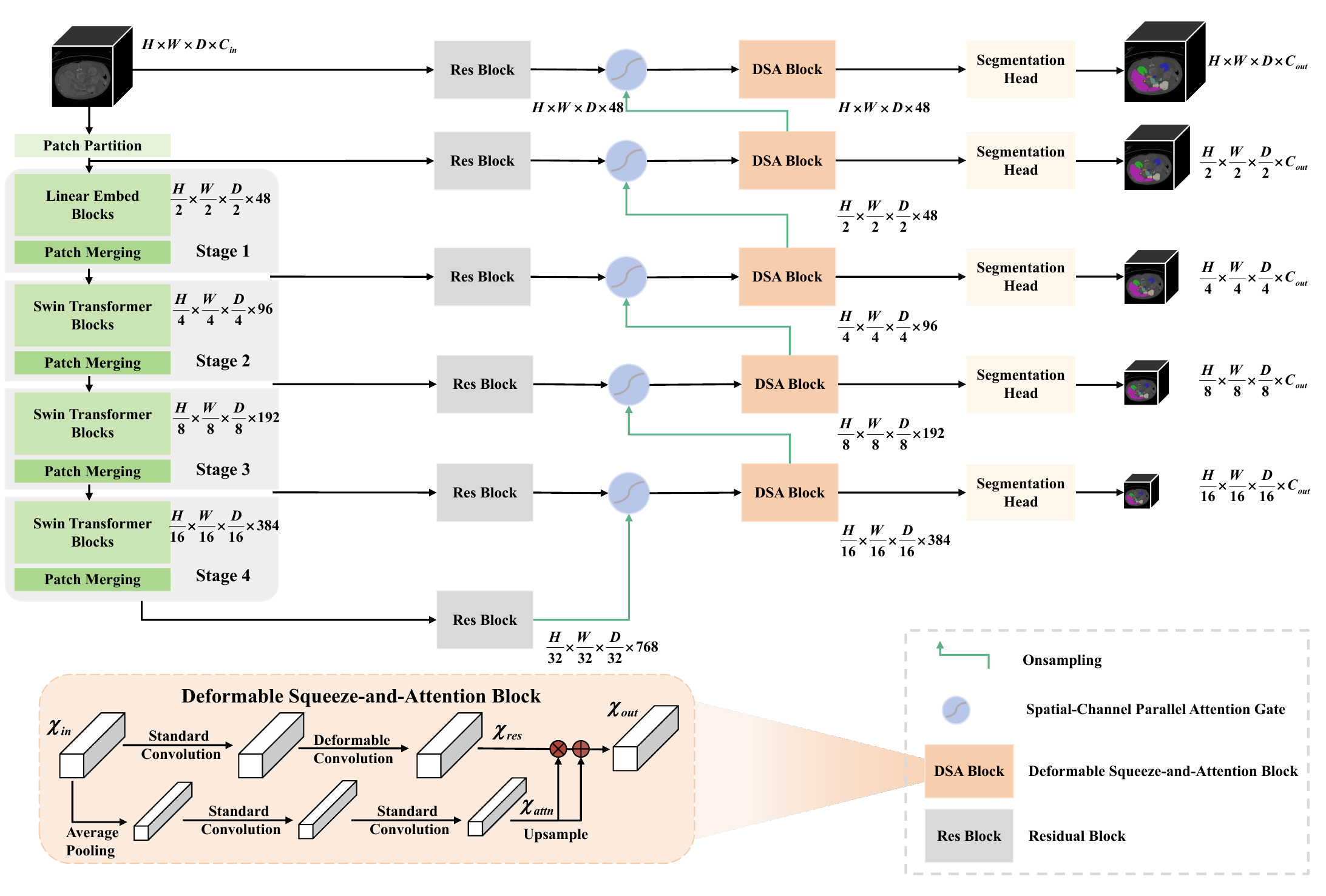}  
    \caption{The overall architecture of our \textbf{Swin DER}. Our design focuses on the decoder side, the feature maps output by the second set of encoder (Residual Block) are reweighted in both spatial and channel dimensions through the \textbf{spatial-channel parallel attention gate} and concatenated in the channel dimension with decoder feature maps upsampled by \textbf{Onsampling} algorithm. Subsequently, the Deformable Squeeze-and-Attention (\textbf{DSA}) Block further integrates these features and utilizes the combined features to learn segmentation details.
}
    \label{network}
\end{figure}

The overall architecture of Swin DER is illustrated in Figure \ref{network}. The network maintains an encoder-decoder structure similar to U-Net\cite{UNet}. The encoder part of Swin DER is consistent with that of Swin UNETR\cite{swinunetr}, comprising two sets based on different architectures: Transformer (Swin Transformer Block) and CNN (Residual Block). The Swin Transformer Block employs a sliding window self-attention mechanism to effectively integrate global information, while the Residual Block captures detailed and local structural information on this basis.

Our design focuses on the decoder side of the network. The decoder utilizes the Offset neighborhood weighted upsampling (Onsampling) algorithm for upsampling, which imparts learnable characteristics to the interpolation algorithm, enhancing its flexibility and effectively avoiding the checkerboard artifacts produced by transposed convolution. The feature maps outputted by encoder are not directly concatenated with the decoder features via skip connections. Instead, they first pass through the spatial-channel parallel attention gate (SCP AG). Under the guidance of decoder features, this process suppresses irrelevant parts between features in both spatial and channel dimensions. The concatenated features are finally sent to the Deformable Squeeze-and-Attention (DSA) Block for feature fusion and detail information learning. Additionally, we have introduced deep supervision during the training process, the outputs of different scales produced by each level of decoders after passing through the segmentation head are compared with their corresponding ground truth segmentation maps (after appropriate scaling) to calculate the loss.

The following sections will introduce these decoder components in detail.

\begin{figure}[!t]
    \centering
    \includegraphics[width=0.75\columnwidth]{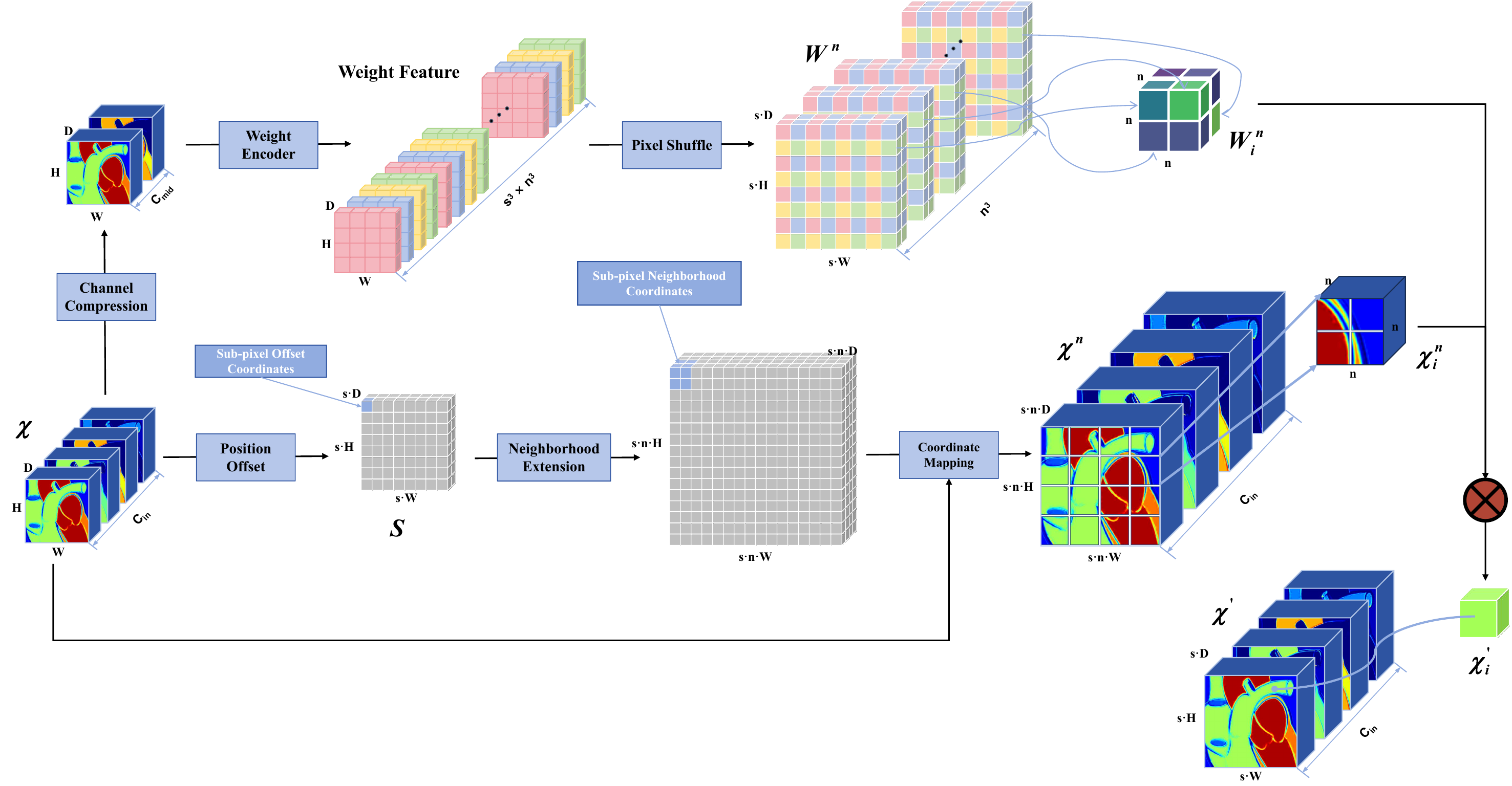}  
    \caption{The computational process of Onsampling. Onsampling adds suitable offsets to the mapping coordinates in trilinear interpolation, enabling the learnability of sub-pixel positions. It obtains neighborhood weights through convolution, making the interpolation weights learnable. Ultimately, these weights are applied to the neighborhood pixels of the offset sub-pixels, realizing a dynamic interpolation algorithm.
}
    \label{onsampling}
\end{figure}

\subsection{Onsampling}
Due to the fixed rules followed by interpolation algorithms in interpolating upsampled values, to increase flexibility, convolution-based upsampling algorithms such as transposed convolution\cite{transconv} and sub-pixel convolution\cite{patil2013medical5} have been introduced. However, these algorithms often come with certain flaws, like the checkerboard artifacts\cite{checkerboard} in transposed convolution. More importantly, we argue that in convolution-based upsampling algorithms, each output pixel is the weighted sum of pixels at corresponding positions across multiple channels. This design causes the output of each channel to be influenced not only by its own channel input but also by inputs from other channels, thereby affecting the clarity and distinctiveness of features in each channel.

Therefore, starting from interpolation algorithms, we designed a learnable interpolation algorithm named \textbf{O}ffset coordinate \textbf{n}eighborhood weighted up\textbf{sampling} (\textbf{Onsampling}). In traditional interpolation algorithms performing $2$-fold upsampling, the sub-pixel positions are situated at the midpoint of the grid formed by the original pixels, and the value of a sub-pixel is derived through the weighted average of neighboring original pixel values, both the positions of sub-pixels and the weights of neighboring pixels are fixed and non-learnable. Onsampling introduces learnability to these two aspects, as illustrated in Figure \ref{onsampling}.

For an input feature $\chi$ of size $H\times W\times D\times C_{in} $, when performing $s$-fold upsampling, the size of the corresponding original sub-pixel coordinate grid $G$ is $s\cdot H\times s\cdot W\times s\cdot D\times 3$. Onsampling adds a learnable offset $O$ to the original sub-pixel coordinates, calculated as follows:
\begin{equation}
    O=0.5\cdot Sigmoid(Conv_{1} (x))\cdot Conv_{2} (x)
\end{equation}
The final sub-pixel coordinate grid $S$ then becomes:
\begin{equation}
    S=O+G
\end{equation}
enabling dynamic changes in sub-pixel positions.

To learn the weights of the neighboring $n$ original pixels of the sub-pixels, onsampling initially applies a $1\times 1\times 1$  convolution to compress the channels to $C_{mid} $. Subsequently, the compressed feature map undergoes encoding to produce the weight feature with dimensions $H\times W\times D\times C_{w} $, where $C_{w}=s^{3} \times  n^{3}$. After pixel shuffling and channel-wise softmax normalization, the final neighborhood weight map $W^{n} $ is obtained.

By expanding the final sub-pixel coordinate grid to its $n$-neighborhood and mapping the input features, a neighborhood feature map $\chi^{n}$ is obtained. Multiplying and summing the $i$-th group of neighborhood features $\chi_{i}^{n} $ with its corresponding neighborhood weights $W_{i}^{n} $ produces the value of the $i$-th pixel value in the output feature map $\chi_{i}^{'}$.

\begin{figure}[!t]
    \centering
    \includegraphics[width=0.75\columnwidth]{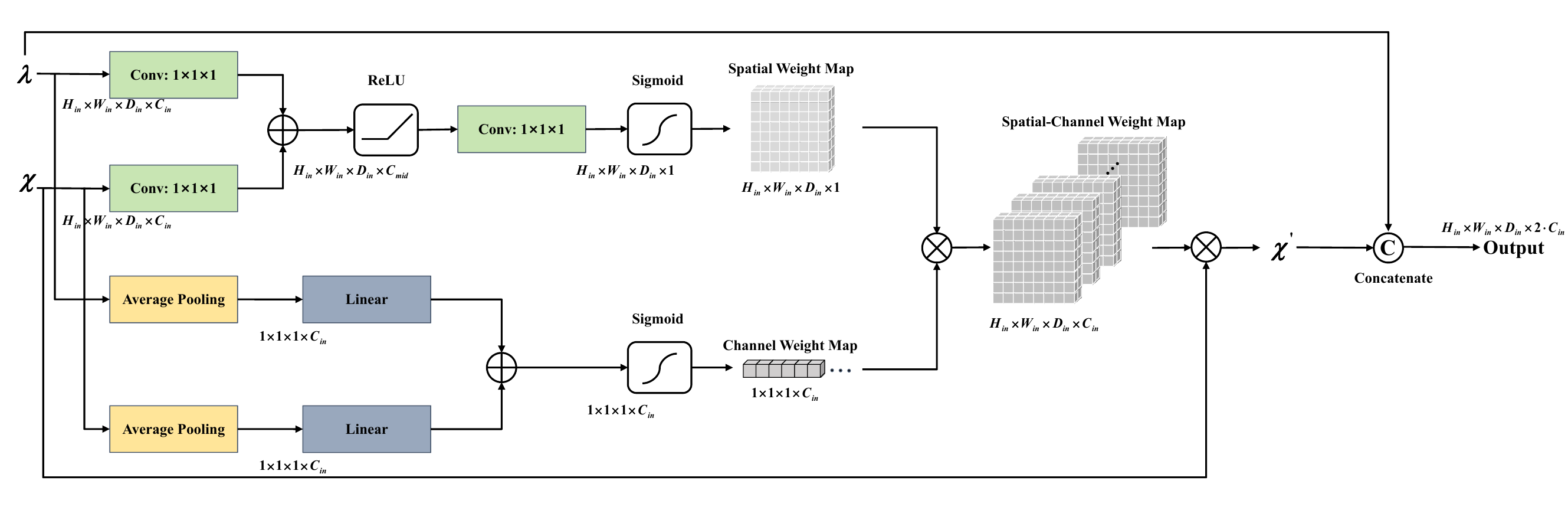} 
    \caption{The schematic diagram of the spatial-channel parallel attention gate. Spatial-Channel Parallel Attention Gate computes weight maps separately in the spatial and channel dimensions, then merges these two-dimensional weight maps to form the spatial-channel weight map. Spatial-channel weight map is used to adjust the significance of each position and channel in the encoder feature map, thereby enhancing the model's focus on important features and suppressing irrelevant information.
}
    \label{SCP_AG}
\end{figure}

\subsection{Spatial-Channel Parallel Attention Gate}
 Spatial-Channel Parallel Attention Gate dynamically adjusts the importance weights of encoder features in both the channel and spatial dimensions by learning the contextual information of the input features, as shown in Figure \ref{SCP_AG}.

In the spatial dimension, encoder features $\chi$ and decoder features $\lambda $ undergo $1\times 1\times 1$ convolution and are then added to generate spatial relevance features. After introducing non-linearity with $ReLU$ activation, these spatial relevance features are compressed across channels through convolution and normalized to a range between 0 and by the $sigmoid$ function, ultimately generating the spatial weight map $W_{S}$ of size $H_{in}\times W_{in}  \times D_{in} \times 1$. The process is as follows:

\begin{equation}
    W_{S} =Sigmoid(Conv(ReLU(Conv_{\chi }(\chi )+Conv_{\lambda }(\lambda ) )))
\end{equation}

In the channel dimension, $\chi$ and $\lambda$ are processed with average pooling, reducing their dimensions to $1\times 1  \times 1 \times C_{in}$ to obtain a global feature representation. The global features are processed through linear mapping and then added together to obtain channel relevance features. Finally, the channel relevance features are activated by the $sigmoid$ function to create the channel weight map $W_{C}$. The calculation formula is as follows:

\begin{equation}
    W_{C} = Sigmoid(Linear_{\chi }(Avg(\chi ))+ Linear_{\lambda  }(Avg(\lambda  )))
\end{equation}

$W_{S}$ and $W_{C}$ independently capture the spatial and channel correlation information within the input feature map. These two weight maps are generated in parallel and integrated together through element-wise multiplication to form the spatial-channel weight map $W_{SC}$:

\begin{equation}
    W_{SC} = W_{S}\otimes W_{c}
\end{equation}
where $\otimes$ denotes element-wise multiplication.

Finally, the encoder feature maps are multiplied by the spatial-channel weight map to enhance the relevant features between the encoder and decoder. Through this mechanism, the spatial-channel parallel attention gate effectively minimizes the semantic gap between features.

\subsection{Deformable Squeeze-and-Attention Block}
The decoder is responsible for integrating features concatenated via skip connections and utilizing these concatenated features to refine segmentation details. Detail features are characterized by locality and irregular shapes. Deformable convolution\cite{deformable} allows for the dynamic adjustment of the shape of convolution kernel to better adapt to the local features of the image, making it highly suitable for learning details.

The output $y(p_{0})$ at position $p_{0}$ in standard convolution can be represented as:
\begin{equation}
    y(p_{0}) = \sum_{p_{n} \subseteq R} w(p_{n}) \cdot x(p_{0} + p_{n})
\end{equation}
where $x$ is the input feature map, $w$ represents the convolution weights, $R$ denotes the convolutional receptive field, and $p_{n}$ represents the relative positions within the convolution kernel. Deformable convolution dynamically computes offsets $\bigtriangleup p_{n}$ for each position $p_{0}$ based on the input feature map through an additional convolutional network, making the output at position \(p_{0}\) become:
\begin{equation}
    y(p_{0})=\sum_{p_{n}\subseteq R }^{} w(p_{n} )\cdot x(p_{_{0} }+p_{n}+\bigtriangleup p_{n})
\end{equation}
The introduction of the $\bigtriangleup p_{n}$ allows the convolution to transcend the constraints of a regular grid shape.

Based on deformable convolution, we propose the Deformable Squeeze-and-Attention (DSA) Block, as illustrated in Figure \ref{network}. In the DSA Block, the output feature map $\chi_{in} $ from the spatial-channel parallel attention gate first undergoes feature fusion through standard convolution. Subsequently, the block employs deformable convolution to learn the segmentation detail information of the fused features. This process can be represented as follows:
\begin{equation}
    \chi_{res}=DefConv(Conv(\chi_{in}))
\end{equation} 
where $DefConv(\cdot )$ and $Conv(\cdot )$ respectively denote deformable convolution and standard convolution.

Simultaneously, $\chi_{in} $ undergoes incomplete average pooling and two consecutive convolutions, finally producing the attention map $\chi_{attn}$ through upsampling:
\begin{equation}
    \chi_{attn}=Upsample(Conv_{2}(Conv_{1}(Avg(\chi_{in}))))
\end{equation}

By multiplication with $\chi_{attn}$, important parts in $\chi_{res}$ are highlighted. After reweighting, $\chi_{res}$ is combined with $\chi_{attn}$ through a residual connection, producing the output $\chi_{out}$. The specific process is detailed as follows:
\begin{equation}
    \chi_{out}=(\chi_{attn}\otimes \chi_{in})\oplus\chi_{attn}
\end{equation} 
where $\otimes$ represents element-wise multiplication, and $\oplus$ denotes the residual connection.

\subsection{Loss function}
We train our networks with a combination of dice\cite{diceloss} and cross-entropy loss, the total loss during the training phase can be formulated as follows:
\begin{equation}
    L_{total} =w_{1} L_{1}+ w_{2} L_{2} + w_{3} L_{3}+ w_{4} L_{4}+ w_{5} L_{5}
\end{equation}
where $L_{i}$, $i\in \left \{1,2,3,4,5  \right \}$ represents the loss of the decoder at the i-th layer. When i equals 1, it represents the topmost decoder layer. Here, $w_{i}$ denotes the weight of loss for the i-th layer of the encoder, the calculation formula is:
\begin{equation}
    w_{i} =\frac{\frac{1}{2^{i-1}}}{\sum_{m=0}^{5}\frac{1}{2^{m}}} 
\end{equation}
The loss for each decoder layer comprises dice loss and cross-entropy loss:
\begin{equation}
    L=L_{dice} + L_{CE}
\end{equation}
The computation formulas for dice loss and cross-entropy loss are as follows:
\begin{equation}
    L_{dice}=1-\frac{2\sum_{c=1}^{C}\sum_{i=1}^{N}g_{i}^{c}s_{i}^{c}}{\sum_{c=1}^{C}\sum_{i=1}^{N}g_{i}^{c}+\sum_{c=1}^{C}\sum_{i=1}^{N}s_{i}^{c}} 
\end{equation}
\begin{equation}
    L_{CE}=- \frac{1}{N} \sum_{c=1}^{C} \sum_{i=1}^{N} g_{i}^{c} \log_{}{s_{i}^{c}} 
\end{equation}
where C represents the number of categories and N represents the number of voxels in 
each category. $g_{i}^{c}$ is the ground truth binary indicator of class label c of voxel i, and $g_{i}^{c}$ is the corresponding segmentation prediction.

\section{Experiments}
\subsection{Datasets}
To validate the effectiveness of our method, we conducted experiments on Synapse multiorgan segmentation\cite{BTCV} and brain tumor segmentation task in Medical Segmentation Decathlon (MSD)\cite{decathlon}. These datasets encompass different imaging modalities and segmentation tasks, providing a comprehensive evaluation of our model.

\subsubsection{Synapse} Dataset comprises abdominal CT scans from 30 subjects with manual annotations conducted under the supervision of radiologists from Vanderbilt University Medical Center, covering 8 distinct organs: spleen, right kidney, left kidney, gallbladder, liver, stomach, aorta, and pancreas. Each CT scan consists of 80 to 225 slices, and each slice having 512$\times$512 pixels with a thickness varying from 1 to 6mm. Following the data split in \cite{transunet}, we select 18 samples for training our model and evaluated it on the remaining 12 samples.

\subsubsection{MSD BraTS task} This task is a part of the Medical Segmentation Decathlon (MSD)\cite{decathlon}, focusing on the segmentation of brain tumors. It comprises 484 MRI images, each including four channels: FLAIR, T1w, T1gd, and T2w. The goal is to accurately label three sub-regions of the tumor: edema (ED), enhancing tumor (ET), and non-enhancing tumor (NET). The data originate from the Brain Tumor Segmentation (BraTS) challenges\cite{Brats} in 2016 and 2017, where the complexity and heterogeneity of the segmentation targets pose significant challenges. Following the data split of UNETR\cite{unetr}, the data is divided into 80\% for training, 15\% for validation, and 5\% for testing. Moreover, to maintain consistency with reported results of other models, We display the segmentation result based on regions, which includes the whole tumor (WT), enhancing tumor (ET), and tumor core (TC). WT encompasses the entire volume of all tumor regions, while TC specifies the central area, excluding the edematous parts of the tumor.

\subsection{Metrics}
We have employed a comprehensive set of two evaluation metrics to rigorously assess the effectiveness of the methodology. These metrics consist of the Dice coefficient, utilized to quantitatively gauge the degree of similarity between the predicted segmentation and the ground truth segmentation. A value converging towards 1 signifies a higher degree of segmentation accuracy. Additionally, we have incorporated the Hausdorff 95 distance, a metric tailored to quantitatively capture the maximum spatial separation between the predicted segmentation and the ground truth. This parameter provides a robust evaluation of the alignment and coherence of segmentation boundaries.The expressions for the two evaluation metrics are provided below:
\begin{equation}
\textrm{Dice} = \frac{2\sum_{i=1}^{I} Y_{i}\hat{Y}_{i} }{\sum_{i=1}^{I}Y_{i}+ \sum_{i=1}^{I}\hat{Y}_{i}},
\label{eq:dice_score}
\end{equation}
\begin{equation}
HD_{95} =\max^{95^{th}} \{{\max _{y' \in Y'} \min _{\Bar{y}' \in \Bar{Y}'} } \|y'-\Bar{y}'\|, 
\max _{\Bar{y}' \in \Bar{Y}'} \min_{y' \in Y'} \|\Bar{y}'-y'\| \}.
\label{eq:hd_score}
\end{equation}
where $Y$ and $\Bar{Y}$ denote the ground truth and prediction of voxel values. $Y'$ and $\hat{Y}'$ denote ground truth and prediction surface point sets. The notation $\max^{95^{th}}(\cdot)$ represents the value obtained by sorting in descending order and selecting the value corresponding to the 95th percentile.

\subsection{Implementation details}
We implement Swin DER in PyTorch\cite{pytorch} 2.0.0 and nnU-Net 2.1.1. All experiments were conducted on NVIDIA GeForce RTX 3090 GPU with 24 GB memory. We follow the default data preprocessing, data augmentation, and training strategies of nnU-Net\cite{nnUNet}. In the data pre-processing stage, we cropped all data to the non-zero regions, then the data will be resampled to the median voxel spacing of the dataset. In the presence of heterogeneous voxel spacings, meaning that the spacing along one axis is three times or more than that of the other axes, the 10 percentile of the spacing will be used as the spatial size for this axis. Finally, the data will be normalized. For CT images, such as synapse, the intensity values of the foreground portion of the dataset are first collected and the entire dataset is normalized by clipping to the [0.5, 99.5] percentiles of these intensity values. Z-score standard normalization\cite{2.5Dzhangchi} then is applied to the data based on the mean and standard deviation of all the collected intensity values. For MRI images, such as MSD BraTS task, or other modalities, individual sample information is collected and z-score normalization is applied to that specific sample. Multiple techniques are employed for data augmentation, including rotation, scaling, Gaussian noise, Gaussian blur, brightness augmentation, contrast adjustment, simulation of low resolution, gamma transformation, and mirror transformation. 

We set the patch size of the Synapse dataset to 128$\times$128$\times$64, while for the MSD BraTS task, the patch size is set to 128$\times$128$\times$128. The batch size for both datasets is 2. 
We utilize the cosine annealing strategy with warm-up to update the learning rate, setting the maximum learning rate to 3e-4:
\begin{equation}
l_{cur} =
\begin{cases} 
l_{initial} \times \frac{E_{cur}}{E_{warmup}}, & \text{if } E_{cur} < E_{warmup} \\
l_{initial} \times \left(1 + \cos \left(\pi \times \frac{E_{cur} - E_{warmup}}{E_{max} - E_{warmup}}\right)\right) / 2, & \text{if } E_{cur} \geq E_{warmup}

\end{cases}
\end{equation}
where $l_{cur}$ denotes the learning rate of the current epoch, $l_{initial}$ is the maximum learning rate after the warm-up phase, $E_{cur}$ denotes the number of current epochs, $E_{warmup}$ is the epoch at which the warm-up ends, and $E_{max}$ denotes the number of training epochs. During the training process, we set $E_{max}$ to 1000 and $E_{warmup}$ to 50. Furthermore, we employ the AdamW optimizer with weight decay of 3e-5 to update gradients. We use the Dice Similarity Coefficient (DSC) and the 95\% Hausdorff Distance (HD95) metrics to evaluate our model.

\subsection{Quantitative results}
To validate the effectiveness of Swin DER on different segmentation tasks, we compared our model with other state-of-the-art methods on the Synapse dataset and MSD BraTs Task. Table.\ref{table:synapse} shows the experimental results of all models on the multi-organ segmentation task. Swin DER achieved the highest average DSC and the lowest average HD95, reaching 86.68\% and 8.64mm, respectively. Additionally, with the design of decoder, Swin DER significantly enhances network performance over Swin UNETR, increasing the average DSC by 4.41\% and reducing the average HD95 by 9.55mm. Compared to second-best method nnFormer, we significantly improved the segmentation performance of the spleen, right kidney, left kidney, gallbaladder, and aorta, with DSC improvements of 1.9\%, 0.85\%, 0.91\%, 4.46\%, and 0.61\%, respectively.

\begin{figure*}[!h]
 \centering
 \includegraphics[width=0.75\textwidth]{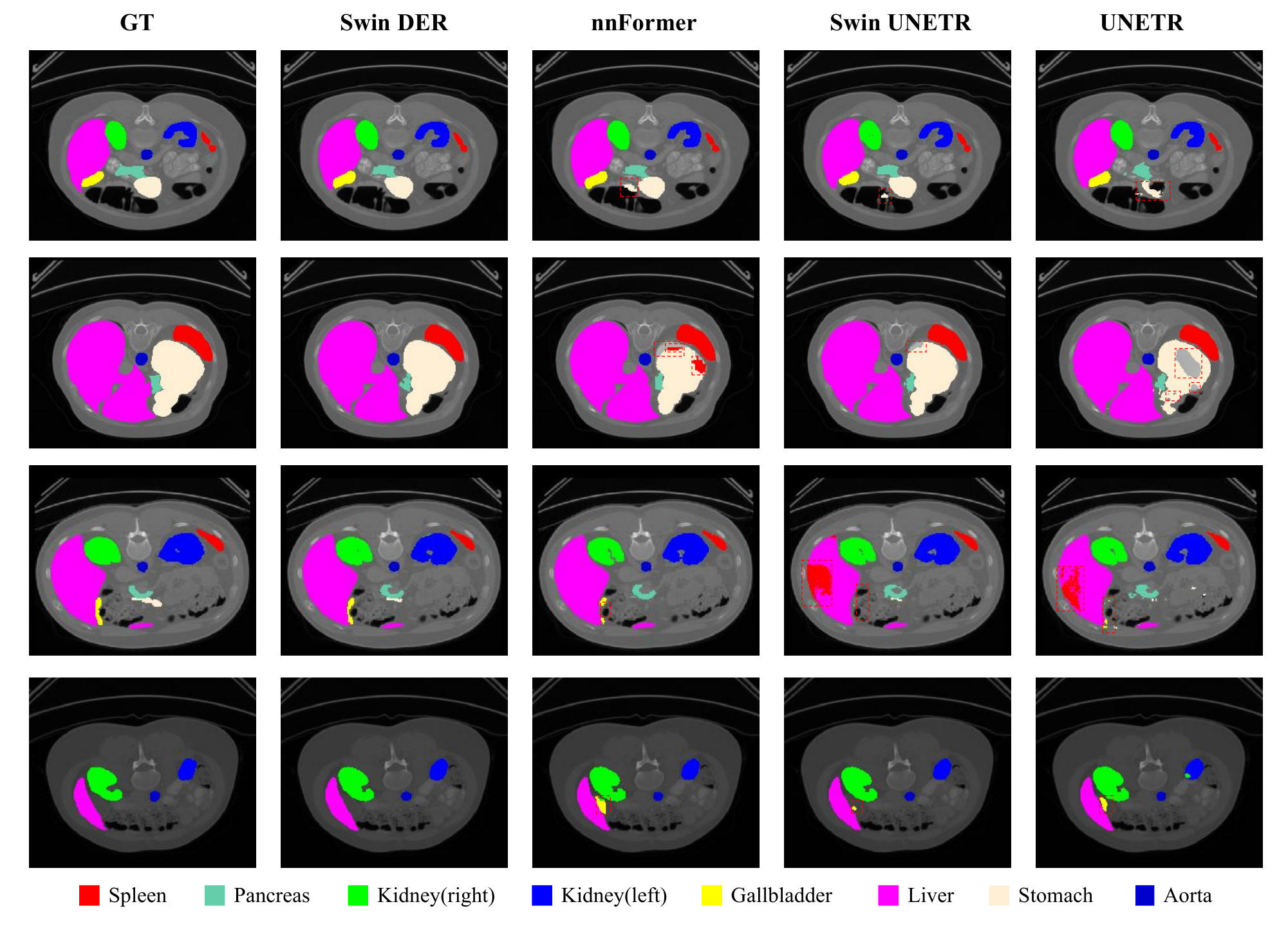}
 \caption{Visualization results of multi-organ segmentation on the Synapse dataset. We primarily compare Swin DER with other transformer-based segmentation models, such as UNETR, Swin UNETR, and nnFormer.}
 \label{fig4}
\end{figure*}

Fig.\ref{fig4} illustrates the qualitative comparison between Swin DER and other methods on the Synapse dataset. As shown in the first row, our method improves the segmentation quality of the stomach. In the second row, nnFormer exhibits over-segmentation of the spleen, which contaminates the segmentation results of the stomach, while both Swin UNETR and UNETR suffer from under-segmentation of the stomach. Swin DER provides a complete segmentation of the stomach and clearly distinguishes the boundary between the stomach and the spleen. In the third row, other methods exhibit varying degrees of under-segmentation of the gallbladder. Additionally, Swin UNETR suffers from severe over-segmentation of the spleen. Only Swin DER provides a complete segmentation of the gallbladder. The fourth row demonstrates Swin DER does not segment a non-existent gallbladder compared to other methods.

\begin{table*}[!t]
    \begin{center}
        \resizebox{\textwidth}{!}{
        \begin{tabular}{l|c c c c c c c c|cc}
        \toprule
            \multirow{2}{*}{Methods}  & \multirow{2}{*}{Spl} &  \multirow{2}{*}{RKid} &  \multirow{2}{*}{ LKid} & \multirow{2}{*}{Gal}  & \multirow{2}{*}{Liv}  & \multirow{2}{*}{Sto} & \multirow{2}{*}{Aor} &  \multirow{2}{*}{Pan} &  \multicolumn{2}{c}{Average} 
            
            \\ \cmidrule{10-11}
             & & & & & & & & & HD95 $\downarrow$ & DSC $\uparrow$ \\
                \midrule
                \midrule
        U-Net~\cite{UNet} & 86.67 & 68.60 & 77.77& 69.72 & 93.43 & 75.58  &  89.07  & 53.98 & - & 76.85 \\
        TransUNet~\cite{transunet} & 85.08 & 77.02 & 81.87 & 63.16 & 94.08 & 75.62 &  87.23  & 55.86 & 31.69 & 77.49 \\
        
        Swin-UNet~\cite{swinunet} & 90.66 &  79.61 & 83.28 & 66.53 & 94.29 & 76.60 & 85.47 &   56.58  & 21.55 & 79.13   \\
        UNETR~\cite{unetr} & 85.00 & 84.52 & 85.60 & 56.30 & 94.57 & 70.46 & 89.80 & 60.47 &  18.59 & 78.35  \\
        MISSFormer~\cite{missformer} & 91.92 &  82.00 & 85.21 &  68.65 & 94.41 & 80.81 & 86.99 & 65.67 & 18.20 & 81.96 \\
        
        Swin UNETR*~\cite{swinunetr} & 88.81 & 85.84 & 86.43 & 65.87 & 95.54 & 75.96 & 89.77 &  69.92 &  15.19 & 82.27 \\
        
        nnFormer~\cite{nnformer} & 90.51 & 86.25 & 86.57 & 70.17 & \textbf{96.84} & \textbf{86.83}  & 92.04 &    \textbf{83.35} &  10.63 & 86.57\\
        \midrule
        \textbf{Swin DER(Ours)} & \textbf{92.41} & \textbf{87.21} & \textbf{87.48} & \textbf{74.63} & 96.24 & 82.43 & \textbf{92.65} & 80.36  &  \textbf{8.64} & \textbf{86.68} \\
        \bottomrule
        \end{tabular}
        }\vspace{-0.5em}
        \caption{\upshape Comparison on the abdominal multi-organ Synapse dataset. We use HD95 and DSC to evaluate the performance of each model. Experimental results of baselines refer to nnFormer\protect\cite{nnformer}. The best results are indicated in bold. * indicates that the baseline is implemented by ourselves. Swin DER achieved the best performance. Abbreviations stand for: Spl: \textit{spleen}, RKid: \textit{right kidney}, LKid: \textit{left kidney}, Gal: \textit{gallbladder}, Liv: \textit{liver}, Sto: \textit{stomach}, Aor: \textit{aorta}, Pan: \textit{pancreas}.}
        \label{table:synapse}
        \end{center}
\vspace{-0.8cm}
\end{table*}

Table.\ref{Brats_table} presents the comparative experimental results on the MSD brain tumor segmentation task, our model achieved the best average DSC and HD95, which are respectively of 86.99\% and 3.65mm. Compared to the baseline model Swin UNETR, the average DSC increased by 1.25\%, and the average HD95 decreased by 0.14 mm. Additionally, Swin DER outperformed the second-ranked method nnFormer in both evaluation metrics across all parts of the brain tumor and in the overall average results. 

\begin{figure}[!h]
    \centering
    \includegraphics[width=0.75\columnwidth]{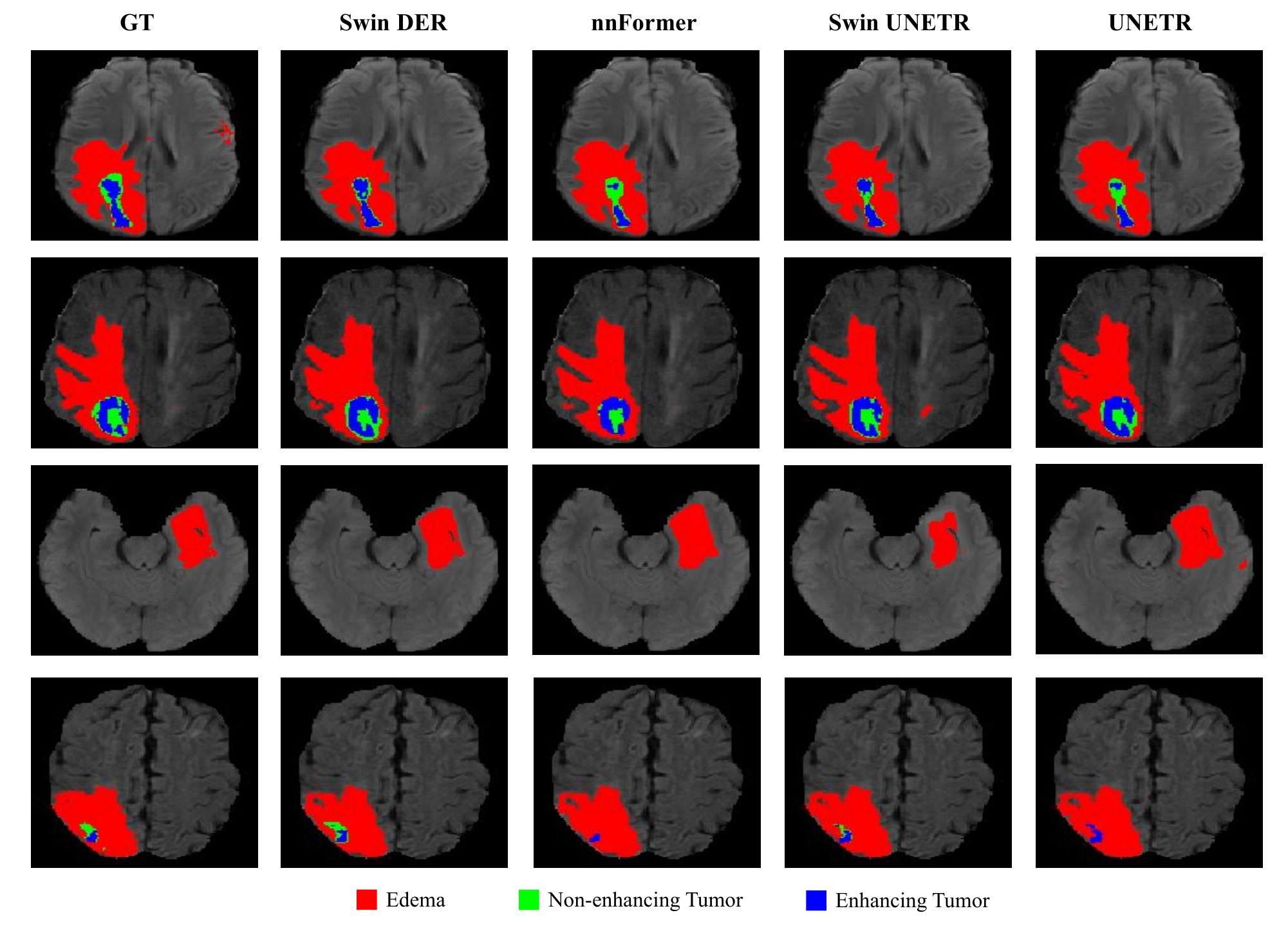}  
    \caption{Brain tumor segmentation visualization. Compared to the current state-of-the-art image segmentation methods, Swin DER achieved the best results.}
    \label{figBraTS}
\end{figure}

Fig.\ref{figBraTS} illustrates the visualization results, in the first line, all methods except Swin DER exhibited under-segmentation of the enhancing tumor. In the second row, both nnFormer and Swin UNETR exhibited over-segmentation of the non-enhancing tumor and under-segmentation of the enhancing tumor. Additionally, compared to UNETR, my method more clearly distinguished the boundaries between the two part. The third row shows the segmentation results of edema using by different methods. It can be observed that nnFormer and UNETR exhibited over-segmentation, while Swin UNETR exhibited under-segmentation. Only Swin DER achieved accurate segmentation of edema. In the fourth row, all methods except Swin DER exhibited under-segmentation of the non-enhancing tumor. Additionally, UNETR also over-segmented the enhancing tumor.

\begin{table*}[!h]
\begin{center}
    \resizebox{1.0\textwidth}{!}{
    \begin{tabular}{l|cc|cc|cc|cc}
    \toprule
       \multirow{2}{*}{Methods} & \multicolumn{2}{c|}{WT} & \multicolumn{2}{c|}{ET} & \multicolumn{2}{c|}{TC} & \multicolumn{2}{c}{Average} \\  \cline{2-9}
       &  HD95 $\downarrow$ & DSC $\uparrow$ & HD95 $\downarrow$ & DSC $\uparrow$ & HD95 $\downarrow$ & DSC $\uparrow$ & HD95 $\downarrow$ & DSC $\uparrow$ \\
       \hline
       \hline
       SETR NUP~\cite{zheng2021rethinking} & 14.419 & 69.7 & 11.72 & 54.4 & 15.19 & 66.9 & 13.78 & 63.7 \\
       SETR PUP~\cite{zheng2021rethinking} & 15.245 & 69.6 & 11.76 & 54.9 & 15.023 & 67.0 & 14.01 & 63.8 \\
       SETR MLA~\cite{zheng2021rethinking} & 15.503 & 69.8 & 10.24 & 55.4 & 14.72 & 66.5 & 13.49 & 63.9 \\
       TransUNet~\cite{transunet} & 14.03 & 70.6 & 10.42 & 54.2 & 14.5 & 68.4 & 12.98 & 64.4 \\
       TransBTS~\cite{wenxuan2021transbts} & 10.03 & 77.9 & 9.97 & 57.4 & 8.95 & 73.5 & 9.65 & 69.6 \\
       CoTr w/o CNN encoder~\cite{xie2021cotr} & 11.49 & 71.2 & 9.59 & 52.3 & 12.58 & 69.8 & 11.22 & 64.4 \\
       CoTr~\cite{xie2021cotr} & 9.20 & 74.6 & 9.45 & 55.7 & 10.45 & 74.8 & 9.70 & 68.3 \\
       UNETR~\cite{unetr} & 8.27 & 78.9 & 9.35 & 58.5 & 8.85 & 76.1 & 8.82 & 71.1 \\
       Swin UNETR*~\cite{swinunetr} & 3.77 & 91.33 & \textbf{2.65} & 80.76 & 4.95 & 85.12 & 3.79 & 85.74 \\
       nnFormer~\cite{nnformer} & 3.80 & 91.3 & 3.87 & 81.8 & 4.49 & 86.0 & 4.05 & 86.4 \\
       \hline
       \textbf{Swin DER(Ours)} & \textbf{3.65} & \textbf{92.27} & 2.94 & \textbf{82.59} & \textbf{4.36} & \textbf{86.10} & \textbf{3.65} & \textbf{86.99} \\
       
    \bottomrule
    \end{tabular}
    }
    \caption{\upshape Comparison on the brain tumor segmentation task in Medical Segmentation Decathlon. We use HD95 and DSC to evaluate the performance of each model. Experimental results of baselines refer to nnFormer\protect\cite{nnformer}. The best results are indicated in bold. * indicates that the baseline is implemented by ourselves. Swin DER achieved the best performance. Abbreviations stand for: WT: \textit{whole tumor}, ET: \textit{enhancing tumor}, TC: \textit{tumor core}.}
    \label{Brats_table}
\end{center}
\end{table*}

\subsection{Ablation study}
\subsubsection{Module ablation experiments}
We conducted ablation experiments on the proposed module using the Synapse dataset, with DSC as the default evaluation metric. Table \ref{tb_abs} displays the detailed experimental results.

The first row in the table represents the baseline model Swin UNETR\cite{swinunetr}. It can be observed that Swin UNETR achieves good segmentation performance on the Synapse dataset, attributed to its complex encoder. However, due to the relatively simple and coarse design of the decoder, its performance on smaller organs like the gallbladder and pancreas is less than ideal with the average DSC lagging behind the current best model nnFormer\cite{nnformer} by 4.3\%, as shown in Table \ref{table:synapse}.

We first replaced the upsampling method in the decoder with Onsampling. Onsampling achieves dynamic interpolation by learning positional offsets and neighborhood weights, effectively improving segmentation performance. As shown in the second row of Table \ref{tb_abs}, compared to using transposed convolution for upsampling, the average DSC increased by 1.7\%, with a significant improvement in the segmentation of the gallbladder and pancreas, where DSC increased by 5.6\% and 4.03\%, respectively.

\begin{table*}[!h]
    \centering
    \resizebox{0.98\textwidth}{!}{
    \begin{tabular}{c|lccccccc|c}
         \toprule
         Models & Spl & RKid & LKid & Gal & Liv & Sto & Aor & Pan & Average\\
         \hline
         \hline
         Encoder + Res Block + Trans Conv (Swin UNETR~\protect\cite{swinunetr}) & 88.81 & 85.84 & 86.43 & 65.87 & 95.54 & 75.96 & 89.77 &  69.92 & 82.27 \\
         \hline
         Encoder + Res Block + Onsampling & 88.01 & 85.84 & 87.14 & 71.47 & 95.16 & 78.47 & 91.69 & 73.95 & 83.97\\
         Encoder + Res Block + SCP AG + Onsampling & \textbf{93.09} & 86.07 & 87.44 & 71.81 & 96.04 & 79.33 & 90.47 & 75.56 & 84.98 \\
         \hline
         Encoder + DSA Block + SCP AG  + Onsampling (Swin DER)  & 92.41 & \textbf{87.21} & \textbf{87.48} & \textbf{74.63} & \textbf{96.24} & \textbf{82.43} & \textbf{92.65} & \textbf{80.36} & \textbf{86.68} \\
         \bottomrule
    \end{tabular}
    }
    \caption{Investigation of the impact of different modules used in Swin DER. Each module in our architecture plays a critical role in enhancing the quality of segmentation. \textbf{Encoder} represents the encoder of Swin UNETR\protect\cite{swinunetr}. \textbf{Res Block} denotes the use of a residual block as the decoder module. \textbf{Trans Conv} represents the use of transposed convolution for upsampling. \textbf{SCP AG} denotes the incorporation of spatial-channel parallel attention gate within the skip connections. \textbf{Onsamping} represents the use of offset coordinate neighborhood weighted upsampling for the upsampling process. \textbf{DSA Block} indicates the use of Deformable Squeeze-and-Attention as the decoder module.} 
    \label{tb_abs}
\end{table*}

Next, we introduced the spatial-channel parallel attention gate, which learns the weights of encoder features within the skip connections in the channel and spatial dimensions. As shown in the third row of Table \ref{tb_abs}, this operation further improved overall performance by 1.01 percents. This indicates that the spatial-channel parallel attention gate effectively bridges the semantic gap between encoder and decoder features, significantly enhancing the skip connections.

Finally, we replaced the feature extraction module on the decoder side with the DSA Block, which further enhanced the performance of model, increasing the average DSC by an additional 1.7\%. As shown in Table \ref{table:synapse}, combining these three modules to specifically improve various parts of the decoder enabled Swin DER to surpass nnFormer, establishing it as the new state-of-the-art model.

\subsubsection{Comparison of different upsampling methods}

The quantitative experiment results of different upsampling methods are shown in Table \ref{tb_upsampling}. Onsampling achieved the highest average DSC of 86.68\% on the Synapes dataset. 

Trilinear Interpolation follows fixed rules, which limits its ability to adapt to the complexity and variability of medical images. As a result, it achieved the lowest average DSC of 85.06\%. Transposed convolution performs upsampling by learning the parameters of convolutional kernels, allowing it to automatically adjust these parameters during training to suit specific tasks. Compared to trilinear interpolation, the use of transposed convolution increased the average DSC by 0.69 percents. However, if not designed properly, it may introduce checkerboard artifacts during the upsampling process. Sub-pixel convolution performs convolution operations on low-resolution feature maps to generate multi-channel feature maps, and then uses pixel shuffle to rearrange these channels into a higher-resolution space. The experimental results in Table \ref{tb_upsampling} shows that there is no significant performance improvement compared to transposed convolution. Additionally, Shi et al.\cite{shi2016deconvolution} have theoretically demonstrated that sub-pixel convolution and transposed convolution do not have substantial differences in their algorithmic essence.

Compared to transposed convolution and sub-pixel convolution, onsampling does not place learnable parameters on the convolution. Instead, it learns the positional offsets and weights of the interpolation reference points, endowing the interpolation algorithm with learnable capabilities and overcoming the shortcomings of convolution-based upsampling methods.

\begin{table}[!h]
	\centering
       \resizebox{0.98\textwidth}{!}{
	\begin{tabular}{l|cccccccc|c}
		\toprule
		Method & Spl & RKid & LKid & Gal & Liv & Sto & Aor & Pan & Average\\
         \hline
         \hline
		Trilinear Interpolation
		& 87.08 & \textbf{87.36} & 87.08 & 70.59 & 95.48 & 82.08 & 91.44 & 79.34 & 85.06 \\
		Transposed Convolution
		& 87.92 & 87.23 & 87.23 & 73.11 & 95.33 & 82.17 & 92.08 & \textbf{80.93} & 85.75 \\
		Sub-pixel Convolution
		& 88.45 & 86.71 & 87.42 & 73.70 & 95.85 & \textbf{83.31} & 92.09 & 79.01 & 85.82 \\
         \hline
		Onsampling
		& \textbf{92.41} & 87.21 & \textbf{87.48} & \textbf{74.63} & \textbf{96.24} & 82.43 & \textbf{92.65} & 80.36 & \textbf{86.68} \\
		\bottomrule
	\end{tabular}
 }
	\vspace{0.1em}
	\caption{\upshape Ablation experiments on upsampling methods.
	}
	\label{tb_upsampling}
	\vspace{-0.1cm}
\end{table}

\subsubsection{Comparison of attention gate}

The relevant experimental results of the attention gate are shown in Table \ref{tb_skip_connection}. When the encoder and decoder feature maps are directly concatenated without using the AG, the semantic gap between the low-level detail information of the encoder features and the high-level semantic information of the decoder features negatively impacts the segmentation performance, resulting in an average DSC of only 85.56\%. The AG reweights the encoder features spatially based on the decoder features, partially bridging the semantic gap between them, which improves the average DSC by 0.2 percents. SCP AG further learns the weights in the channel dimension and combining them with spatial weights to reweight the decoder features across both dimensions, achieving the best average DSC of 86.68\%.

\begin{table}[!h]
	\centering
       \resizebox{0.98\textwidth}{!}{
	\begin{tabular}{l|cccccccc|c}
		\toprule
		Method & Spl & RKid & LKid & Gal & Liv & Sto & Aor & Pan & Average\\
         \hline
         \hline
		No Attention Gate
		& 87.22 & \textbf{87.39} & 87.18 & 74.45 & 95.42 & 82.36 & 91.59 & 78.84 & 85.56 \\
		Attention Gate
		& 92.21 & 86.37 & 87.15 & 73.19 & 95.31 & 82.13 & 92.23 & 77.49 & 85.76 \\
         \hline
		 Spatial-Channel Parallel Attention Gate
		& \textbf{92.41} & 87.21 & \textbf{87.48} & \textbf{74.63} & \textbf{96.24} & \textbf{82.43} & \textbf{92.65} & \textbf{80.36} & \textbf{86.68} \\
		\bottomrule
	\end{tabular}
 }
	\vspace{0.1em}
	\caption{\upshape Ablation experiments on the attention gate of skip connection.
	}
	\label{tb_skip_connection}
	\vspace{-0.1cm}
\end{table}

\subsubsection{Comparison of feature extraction module of decoder}

The ablation experiment results for the decoder feature extraction module are shown in Table \ref{tb_eature_extraction_module}, where the Basic Block represents two consecutive convolution layers. The residual block, through additional connections, reduces the risk of overfitting and enhances the generalization ability of model. Compared to the basic block, it improved the average DSC by 0.24\%. Deformable convolution\cite{deformable} can adaptively adjust sampling positions, allowing it to better preserve and utilize important information when dealing with complex spatial structures. The results in the fourth and fifth rows of the Table \ref{tb_eature_extraction_module} show that after replacing the second convolution layer in both the residual block and the basic block with deformable convolution, the average DSC increased by 0.31 and 0.12 percentage points respectively. The DSA Block introduces attention mechanism into the additional connection of the residual module with deformable convolution, assigning higher weight to important features and providing guidance for the feature extraction module, achieving the best segmentation results.

\begin{table}[!t]
	\centering
       \resizebox{0.98\textwidth}{!}{
	\begin{tabular}{l|cccccccc|c}
		\toprule
		Method & Spl & RKid & LKid & Gal & Liv & Sto & Aor & Pan & Average\\
         \hline
         \hline
	   Basic Block
		& 89.54 & 87.19 & 87.51 & 67.85 & 95.57 & 81.79 & 91.40 & 77.04 & 84.74 \\
	   Res Block
            & \textbf{93.09} & 86.07 & 87.44 & 71.81 & 96.04 & 79.33 & 90.47 & 75.56 & 84.98 \\
          Basic Block with Deformable Conv
		& 87.76 & \textbf{87.38} & \textbf{87.60} & 70.86 & 95.59 & 80.02 & 92.18 & 77.52 & 84.86 \\
	   Res Block with Deformable Conv
            & 87.67 & 87.07 & 87.27 & 72.82 & 95.35 & 81.54 & 90.87 & 79.72 & 85.29 \\
         \hline
		 DSA Block
		& 92.41 & 87.21 & 87.48 & \textbf{74.63} & \textbf{96.24} & \textbf{82.43} & \textbf{92.65} & \textbf{80.36} & \textbf{86.68} \\
		\bottomrule
	\end{tabular}
 }
	\vspace{0.1em}
	\caption{\upshape Ablation experiments on the feature extraction module of the decoder.
	}
	\label{tb_eature_extraction_module}
	\vspace{-0.1cm}
\end{table}

\section{Conclusion}
In this paper, we propose a network focused on decoder design, called Swin DER. Swin DER enhances the upsampling process, skip connection, and decoder feature extraction module. We design a novel and efficient upsampling method called Onsampling, which improves the flexibility and scalability of the interpolation algorithm by learning the positional offsets and weights of neighboring pixels. Onsampling also avoids the drawbacks of convolution-based upsampling methods. For skip connection, we introduce a spatial-channel parallel attention gate, which weights the encoder features across both spatial and channel dimensions. This helps bridge the significant semantic gap between the encoder and decoder features. Additionally, we introduce deformable convolution and attention mechanism into the feature extraction module. This combination allows the decoder to dynamically adjust its receptive field while giving more focus to important features, thereby enhancing the feature extraction and learning capabilities of decoder. 

Experimental results show that on both the Synapse and the MSD brain tumor segmentation task, Swin DER outperforms other state-of-the-art methods, particularly those based on transformer. This proves the superiority of our method. ore importantly, it demonstrates that optimizing the decoder can further enhance segmentation capability of model.


\section{Acknowledge}
This work was supported by National Natural Science Foundation of China under Grant 61301253 and the Major Scientific and Technological Innovation Project in Shandong Province under Grant 2021CXG010506 and 2022CXG010504;"New Universities 20 items" Funding Project of Jinan under Grant 2021GXRC108 and 2021GXRC024.

\section*{References}
\bibliographystyle{jphysicsB}  
\bibliography{mylib}

@article{overview1,
  title={Current and emerging trends in medical image segmentation with deep learning},
  author={Conze, Pierre-Henri and Andrade-Miranda, Gustavo and Singh, Vivek Kumar and Jaouen, Vincent and Visvikis, Dimitris},
  journal={IEEE Transactions on Radiation and Plasma Medical Sciences},
  year={2023},
  publisher={IEEE}
}

@article{overview2,
  title={A review of deep-learning-based medical image segmentation methods},
  author={Liu, Xiangbin and Song, Liping and Liu, Shuai and Zhang, Yudong},
  journal={Sustainability},
  volume={13},
  number={3},
  pages={1224},
  year={2021},
  publisher={MDPI}
}

@article{overview3,
  title={Medical image segmentation using deep learning: A survey},
  author={Wang, Risheng and Lei, Tao and Cui, Ruixia and Zhang, Bingtao and Meng, Hongying and Nandi, Asoke K},
  journal={IET Image Processing},
  volume={16},
  number={5},
  pages={1243--1267},
  year={2022},
  publisher={Wiley Online Library}
}

@article{overview4,
  title={Segment anything model for medical image segmentation: Current applications and future directions},
  author={Zhang, Yichi and Shen, Zhenrong and Jiao, Rushi},
  journal={Computers in Biology and Medicine},
  pages={108238},
  year={2024},
  publisher={Elsevier}
}

@article{decathlon,
  title={The medical segmentation decathlon},
  author={Antonelli, Michela and Reinke, Annika and Bakas, Spyridon and Farahani, Keyvan and Kopp-Schneider, Annette and Landman, Bennett A and Litjens, Geert and Menze, Bjoern and Ronneberger, Olaf and Summers, Ronald M and others},
  journal={Nature communications},
  volume={13},
  number={1},
  pages={4128},
  year={2022},
  publisher={Nature Publishing Group UK London}
}

@inproceedings{wenxuan2021transbts,
  title={Transbts: Multimodal brain tumor segmentation using transformer},
  author={Wenxuan, Wang and Chen, Chen and Meng, Ding and Hong, Yu and Sen, Zha and Jiangyun, Li},
  booktitle={International Conference on Medical Image Computing and Computer-Assisted Intervention, Springer},
  pages={109--119},
  year={2021}
}

@inproceedings{xie2021cotr,
  title={Cotr: Efficiently bridging cnn and transformer for 3d medical image segmentation},
  author={Xie, Yutong and Zhang, Jianpeng and Shen, Chunhua and Xia, Yong},
  booktitle={Medical Image Computing and Computer Assisted Intervention--MICCAI 2021: 24th International Conference, Strasbourg, France, September 27--October 1, 2021, Proceedings, Part III 24},
  pages={171--180},
  year={2021},
  organization={Springer}
}

@inproceedings{zheng2021rethinking,
  title={Rethinking semantic segmentation from a sequence-to-sequence perspective with transformers},
  author={Zheng, Sixiao and Lu, Jiachen and Zhao, Hengshuang and Zhu, Xiatian and Luo, Zekun and Wang, Yabiao and Fu, Yanwei and Feng, Jianfeng and Xiang, Tao and Torr, Philip HS and others},
  booktitle={Proceedings of the IEEE/CVF conference on computer vision and pattern recognition},
  pages={6881--6890},
  year={2021}
}

@article{Brats,
  title={The multimodal brain tumor image segmentation benchmark (BRATS)},
  author={Menze, Bjoern H and Jakab, Andras and Bauer, Stefan and Kalpathy-Cramer, Jayashree and Farahani, Keyvan and Kirby, Justin and Burren, Yuliya and Porz, Nicole and Slotboom, Johannes and Wiest, Roland and others},
  journal={IEEE transactions on medical imaging},
  volume={34},
  number={10},
  pages={1993--2024},
  year={2014},
  publisher={IEEE}
}

@article{wasserthal2023totalsegmentator,
  title={Totalsegmentator: Robust segmentation of 104 anatomic structures in ct images},
  author={Wasserthal, Jakob and Breit, Hanns-Christian and Meyer, Manfred T and Pradella, Maurice and Hinck, Daniel and Sauter, Alexander W and Heye, Tobias and Boll, Daniel T and Cyriac, Joshy and Yang, Shan and others},
  journal={Radiology: Artificial Intelligence},
  volume={5},
  number={5},
  year={2023},
  publisher={Radiological Society of North America}
}

@article{overview5,
  title={Medical image segmentation review: The success of u-net},
  author={Azad, Reza and Aghdam, Ehsan Khodapanah and Rauland, Amelie and Jia, Yiwei and Avval, Atlas Haddadi and Bozorgpour, Afshin and Karimijafarbigloo, Sanaz and Cohen, Joseph Paul and Adeli, Ehsan and Merhof, Dorit},
  journal={arXiv preprint arXiv:2211.14830},
  year={2022}
}

@article{overview6,
  title={U-net and its variants for medical image segmentation: A review of theory and applications},
  author={Siddique, Nahian and Paheding, Sidike and Elkin, Colin P and Devabhaktuni, Vijay},
  journal={Ieee Access},
  volume={9},
  pages={82031--82057},
  year={2021},
  publisher={IEEE}
}

@article{overview7,
  title={Transformers in medical imaging: A survey},
  author={Shamshad, Fahad and Khan, Salman and Zamir, Syed Waqas and Khan, Muhammad Haris and Hayat, Munawar and Khan, Fahad Shahbaz and Fu, Huazhu},
  journal={Medical Image Analysis},
  pages={102802},
  year={2023},
  publisher={Elsevier}
}

@article{transformer,
  title={An image is worth 16x16 words: Transformers for image recognition at scale},
  author={Dosovitskiy, Alexey and Beyer, Lucas and Kolesnikov, Alexander and Weissenborn, Dirk and Zhai, Xiaohua and Unterthiner, Thomas and Dehghani, Mostafa and Minderer, Matthias and Heigold, Georg and Gelly, Sylvain and others},
  journal={arXiv preprint arXiv:2010.11929},
  year={2020}
}

@inproceedings{swinunet,
  title={Swin-unet: Unet-like pure transformer for medical image segmentation},
  author={Cao, Hu and Wang, Yueyue and Chen, Joy and Jiang, Dongsheng and Zhang, Xiaopeng and Tian, Qi and Wang, Manning},
  booktitle={European conference on computer vision},
  pages={205--218},
  year={2022},
  organization={Springer}
}

@inproceedings{UNet3+,
  title={Unet 3+: A full-scale connected unet for medical image segmentation},
  author={Huang, Huimin and Lin, Lanfen and Tong, Ruofeng and Hu, Hongjie and Zhang, Qiaowei and Iwamoto, Yutaro and Han, Xianhua and Chen, Yen-Wei and Wu, Jian},
  booktitle={ICASSP 2020-2020 IEEE international conference on acoustics, speech and signal processing (ICASSP)},
  pages={1055--1059},
  year={2020},
  organization={IEEE}
}

@article{UNet++,
  title={Unet++: Redesigning skip connections to exploit multiscale features in image segmentation},
  author={Zhou, Zongwei and Siddiquee, Md Mahfuzur Rahman and Tajbakhsh, Nima and Liang, Jianming},
  journal={IEEE transactions on medical imaging},
  volume={39},
  number={6},
  pages={1856--1867},
  year={2019},
  publisher={IEEE}
}

@article{Attention_UNet,
  title={Attention u-net: Learning where to look for the pancreas},
  author={Oktay, Ozan and Schlemper, Jo and Folgoc, Loic Le and Lee, Matthew and Heinrich, Mattias and Misawa, Kazunari and Mori, Kensaku and McDonagh, Steven and Hammerla, Nils Y and Kainz, Bernhard and others},
  journal={arXiv preprint arXiv:1804.03999},
  year={2018}
}

@inproceedings{unetr,
  title={Unetr: Transformers for 3d medical image segmentation},
  author={Hatamizadeh, Ali and Tang, Yucheng and Nath, Vishwesh and Yang, Dong and Myronenko, Andriy and Landman, Bennett and Roth, Holger R and Xu, Daguang},
  booktitle={Proceedings of the IEEE/CVF winter conference on applications of computer vision},
  pages={574--584},
  year={2022}
}

@inproceedings{swinunetr,
  title={Swin unetr: Swin transformers for semantic segmentation of brain tumors in mri images},
  author={Hatamizadeh, Ali and Nath, Vishwesh and Tang, Yucheng and Yang, Dong and Roth, Holger R and Xu, Daguang},
  booktitle={International MICCAI Brainlesion Workshop},
  pages={272--284},
  year={2021},
  organization={Springer}
}

@article{shi2016deconvolution,
  title={Is the deconvolution layer the same as a convolutional layer?},
  author={Shi, Wenzhe and Caballero, Jose and Theis, Lucas and Huszar, Ferenc and Aitken, Andrew and Ledig, Christian and Wang, Zehan},
  journal={arXiv preprint arXiv:1609.07009},
  year={2016}
}

@inproceedings{transconv,
  title={Learning deconvolution network for semantic segmentation},
  author={Noh, Hyeonwoo and Hong, Seunghoon and Han, Bohyung},
  booktitle={Proceedings of the IEEE international conference on computer vision},
  pages={1520--1528},
  year={2015}
}

@inproceedings{skipconnection1,
  title={The importance of skip connections in biomedical image segmentation},
  author={Drozdzal, Michal and Vorontsov, Eugene and Chartrand, Gabriel and Kadoury, Samuel and Pal, Chris},
  booktitle={International workshop on deep learning in medical image analysis, international workshop on large-scale annotation of biomedical data and expert label synthesis},
  pages={179--187},
  year={2016},
  organization={Springer}
}

@inproceedings{deformable,
  title={Deformable convolutional networks},
  author={Dai, Jifeng and Qi, Haozhi and Xiong, Yuwen and Li, Yi and Zhang, Guodong and Hu, Han and Wei, Yichen},
  booktitle={Proceedings of the IEEE international conference on computer vision},
  pages={764--773},
  year={2017}
}

@inproceedings{skipconnection2,
  title={Going deeper with convolutions},
  author={Szegedy, Christian and Liu, Wei and Jia, Yangqing and Sermanet, Pierre and Reed, Scott and Anguelov, Dragomir and Erhan, Dumitru and Vanhoucke, Vincent and Rabinovich, Andrew},
  booktitle={Proceedings of the IEEE conference on computer vision and pattern recognition},
  pages={1--9},
  year={2015}
}

@inproceedings{BiO-Net,
  title={BiO-Net: learning recurrent bi-directional connections for encoder-decoder architecture},
  author={Xiang, Tiange and Zhang, Chaoyi and Liu, Dongnan and Song, Yang and Huang, Heng and Cai, Weidong},
  booktitle={Medical Image Computing and Computer Assisted Intervention--MICCAI 2020: 23rd International Conference, Lima, Peru, October 4--8, 2020, Proceedings, Part I 23},
  pages={74--84},
  year={2020},
  organization={Springer}
}

@article{vit,
  title={An image is worth 16x16 words: Transformers for image recognition at scale},
  author={Dosovitskiy, Alexey and Beyer, Lucas and Kolesnikov, Alexander and Weissenborn, Dirk and Zhai, Xiaohua and Unterthiner, Thomas and Dehghani, Mostafa and Minderer, Matthias and Heigold, Georg and Gelly, Sylvain and others},
  journal={arXiv preprint arXiv:2010.11929},
  year={2020}
}

@article{multiresunet,
  title={MultiResUNet: Rethinking the U-Net architecture for multimodal biomedical image segmentation},
  author={Ibtehaz, Nabil and Rahman, M Sohel},
  journal={Neural networks},
  volume={121},
  pages={74--87},
  year={2020},
  publisher={Elsevier}
}

@inproceedings{BTCV,
  title={Miccai multi-atlas labeling beyond the cranial vault--workshop and challenge},
  author={Landman, Bennett and Xu, Zhoubing and Igelsias, J and Styner, Martin and Langerak, T and Klein, Arno},
  booktitle={Proc. MICCAI Multi-Atlas Labeling Beyond Cranial Vault—Workshop Challenge},
  volume={5},
  pages={12},
  year={2015}
}

@inproceedings{UNet,
  title={U-net: Convolutional networks for biomedical image segmentation},
  author={Ronneberger, Olaf and Fischer, Philipp and Brox, Thomas},
  booktitle={Medical Image Computing and Computer-Assisted Intervention--MICCAI 2015: 18th International Conference, Munich, Germany, October 5-9, 2015, Proceedings, Part III 18},
  pages={234--241},
  year={2015},
  organization={Springer}
}

@article{patil2013medical3,
  title={Survey: Interpolation methods in medical image processing},
  author={Lehmann, Thomas Martin and Gonner, Claudia and Spitzer, Klaus},
  journal={IEEE transactions on medical imaging},
  volume={18},
  number={11},
  pages={1049--1075},
  year={1999},
  publisher={IEEE}
}

@inproceedings{patil2013medical5,
  title={Real-time single image and video super-resolution using an efficient sub-pixel convolutional neural network},
  author={Shi, Wenzhe and Caballero, Jose and Husz{\'a}r, Ferenc and Totz, Johannes and Aitken, Andrew P and Bishop, Rob and Rueckert, Daniel and Wang, Zehan},
  booktitle={Proceedings of the IEEE conference on computer vision and pattern recognition},
  pages={1874--1883},
  year={2016}
}

@article{nnUNet,
  title={nnU-Net: a self-configuring method for deep learning-based biomedical image segmentation},
  author={Isensee, Fabian and Jaeger, Paul F and Kohl, Simon AA and Petersen, Jens and Maier-Hein, Klaus H},
  journal={Nature methods},
  volume={18},
  number={2},
  pages={203--211},
  year={2021},
  publisher={Nature Publishing Group}
}

@article{IN,
  title={Instance normalization: The missing ingredient for fast stylization},
  author={Ulyanov, Dmitry and Vedaldi, Andrea and Lempitsky, Victor},
  journal={arXiv preprint arXiv:1607.08022},
  year={2016}
}

@article{transunet,
  title={Transunet: Transformers make strong encoders for medical image segmentation},
  author={Chen, Jieneng and Lu, Yongyi and Yu, Qihang and Luo, Xiangde and Adeli, Ehsan and Wang, Yan and Lu, Le and Yuille, Alan L and Zhou, Yuyin},
  journal={arXiv preprint arXiv:2102.04306},
  year={2021}
}

@article{missformer,
  title={Missformer: An effective medical image segmentation transformer},
  author={Huang, Xiaohong and Deng, Zhifang and Li, Dandan and Yuan, Xueguang},
  journal={arXiv preprint arXiv:2109.07162},
  year={2021}
}

@article{nnformer,
  title={nnformer: Interleaved transformer for volumetric segmentation},
  author={Zhou, Hong-Yu and Guo, Jiansen and Zhang, Yinghao and Yu, Lequan and Wang, Liansheng and Yu, Yizhou},
  journal={arXiv preprint arXiv:2109.03201},
  year={2021}
}

@incollection{pytorch,
title = {PyTorch: An Imperative Style, High-Performance Deep Learning Library},
author = {Paszke, Adam and Gross, Sam and Massa, Francisco and Lerer, Adam and Bradbury, James and Chanan, Gregory and Killeen, Trevor and Lin, Zeming and Gimelshein, Natalia and Antiga, Luca and Desmaison, Alban and Kopf, Andreas and Yang, Edward and DeVito, Zachary and Raison, Martin and Tejani, Alykhan and Chilamkurthy, Sasank and Steiner, Benoit and Fang, Lu and Bai, Junjie and Chintala, Soumith},
booktitle = {Advances in Neural Information Processing Systems 32},
editor = {H. Wallach and H. Larochelle and A. Beygelzimer and F. d\textquotesingle Alch\'{e}-Buc and E. Fox and R. Garnett},
pages = {8024--8035},
year = {2019},
publisher = {Curran Associates, Inc.},
url = {http://papers.neurips.cc/paper/9015-pytorch-an-imperative-style-high-performance-deep-learning-library.pdf}
}

@article{2.5Dzhangchi,
  title={Liver tumor segmentation using 2.5 D UV-Net with multi-scale convolution},
  author={Zhang, Chi and Hua, Qianqian and Chu, Yingying and Wang, Pengwei},
  journal={Computers in Biology and Medicine},
  volume={133},
  pages={104424},
  year={2021},
  publisher={Elsevier}
}

@inproceedings{diceloss,
  title={The importance of skip connections in biomedical image segmentation},
  author={Drozdzal, Michal and Vorontsov, Eugene and Chartrand, Gabriel and Kadoury, Samuel and Pal, Chris},
  booktitle={International Workshop on Deep Learning in Medical Image Analysis, International Workshop on Large-Scale Annotation of Biomedical Data and Expert Label Synthesis},
  pages={179--187},
  year={2016},
  organization={Springer}
}

@article{checkerboard,
  title={Deconvolution and checkerboard artifacts},
  author={Odena, Augustus and Dumoulin, Vincent and Olah, Chris},
  journal={Distill},
  volume={1},
  number={10},
  pages={e3},
  year={2016}
}
\end{document}